\documentstyle[12pt,epsf]{article}

\newcommand{\beq}{\begin{equation}}
\newcommand{\eeq}{\end{equation}}
\newcommand{\beqa}{\begin{eqnarray}}
\newcommand{\eeqa}{\end{eqnarray}}
\newcommand{\no}{\nonumber}
\newcommand{\q}{\quad}
\newcommand{\qq}{\qquad}
\newcommand{\mnod}{\stackrel{\circ}{M}}

\newcommand{\tr}{\mbox{tr}}

\begin{document}

\hfill 

\hfill 

\bigskip\bigskip

\begin{center}

{{\Large\bf Resonances in radiative hyperon decays}}

\end{center}

\vspace{.4in}

\begin{center}
{\large Bu\={g}ra Borasoy\footnote{email: borasoy@het.phast.umass.edu}
 and Barry R. Holstein\footnote{email: holstein@phast.umass.edu }}

\bigskip

\bigskip

Department of Physics and Astronomy\\
University of Massachusetts\\
Amherst, MA 01003, USA\\

\vspace{.2in}

\end{center}

\vspace{.7in}

\thispagestyle{empty} 

\begin{abstract}
The importance of resonances for the radiative hyperon decays is 
examined in the framework of chiral perturbation theory. 
Low lying  baryon resonances are included into the effective theory
and tree contributions to these decays are calculated.
We find significant contributions to both the parity-conserving and
parity-violating decay amplitudes
and a large negative value for the asymmetry parameter in polarized
$\Sigma^+ \rightarrow p \gamma$ is found, in agreement with the
experimental result $\alpha_\gamma^{p\Sigma^+}=-0.76\pm 0.08$.
\end{abstract}

\vfill

\section{Introduction}

The radiative hyperon decays---$\Sigma^+\rightarrow
p\gamma,\Lambda\rightarrow n\gamma,$ {\it etc.}---have been studied both
experimentally and theoretically for over three decades, but still a
number of mysteries still exist.\cite{mrz}  The primary problem has been
and remains to understand the size of the asymmetry parameter in
polarized $\Sigma^+\rightarrow p\gamma$ decay\cite{pdb}
\begin{equation}
\alpha_\gamma=-0.76\pm 0.08\label{eq:aa}
\end{equation}
The difficulty here is associated with the restrictions posed by Hara's
theorem, which requires the vanishing of this asymmetry in the SU(3)
limit.\cite{hara}  The proof of this theorem is easily given.  By
gauge invariance the radiative decay amplitudes must have the form
\begin{equation}
{\rm Amp}(B\rightarrow B'\gamma)=\bar{u}_{B'}(p')
{-i \over2(M_B+M_{B'})}\sigma_{\mu\nu}F^{\mu\nu}
(A_{B'B}+B_{B'B}\gamma_5)u_B(p)\label{eq:bb}
\end{equation}
where $F^{\mu\nu}$ is the electromagnetic field strength tensor, 
$A_{B'B}$ is the parity conserving M1 amplitude and $B_{B'B}$ is
its parity violating E1 counterpart.  Now under U-spin, the
electromagnetic current is a singlet, while the weak $\Delta S=1$
Hamiltonian acts like the $\Delta U_3=-1$ component of a U-spin
vector.  Thus the effective current for this transition transforms
like a CP-even U-spin lowering operator.  This is completely
analogous to the situation in nuclear beta decay involving the isospin
operator, and the same arguments which require the vanishing of
tensor matrix
element---$-i\sigma_{\mu\nu}\gamma_5F^{\mu\nu} $---for the axial current
between isospin analog states such as $n,p$\cite{ht} 
guarantee the vanishing of
the E1 radiative hyperon decay amplitude between states such as
$\Sigma^+,p$ which are members of a common U-spin
multiplet.\footnote{Note that the same argument would apply to the
transition $\Xi^-\rightarrow\Sigma^-\gamma$.  However, there is
presently no experimental information on the asymmetry for this
decay.}  Since the asymmetry parameter is related to the decay amplitudes via 
\begin{equation}
\alpha_\gamma= {2{\rm Re}A_{B'B}B_{B'B}^*\over
|A_{B'B}|^2+|B_{B'B}|^2}
\label{eq:qq}
\end{equation}
the vanishing of the E1 amplitude also guarantees the vanishing of
the asymmetry in contradiction to the near maximal 
number---Eq. \ref {eq:aa}---measured experimentally.  

Of course, in the real world U-spin is broken and one should not be
surprised to find a nonzero value for the asymmetry---what {\it is}
difficult to understand is its size.  Indeed theorists have modelled
this decay for nearly three decades and a definitive explanation for
the $\Sigma^+\rightarrow p\gamma$ asymmetry remains lacking.  This is
certainly {\it not} for lack of trying---indeed many ideas have been
pursued.\cite{mrz}  Early approaches utilized simple baryon pole models with 
weak parity conserving $BB'$ matrix elements determined from 
fits to ordinary hyperon
decay.\cite{pole}  The SU(3) breaking in this case comes from the difference
between experimental $\Sigma^+$ and proton magnetic moments, which
leads to asymmetry estimates in the 10\% range.  Corresponding parity 
violating $BB'$ matrix elements must vanish in the SU(3) limit via the
Lee-Swift theorem.\cite{ls}  However, SU(3) breaking leads to a nonvanishing
asymmetry again only at the $\sim$10\% level.\cite{gh}  Recent work involving
the calculation of chiral loops has also not lead to a resolution, although
slightly larger asymmetries can be accomodated.\cite{N}  In addition, 
over the years there have been claims that Hara's theorem is not to be
trusted.\cite{noh}  However, these have proven to be incorrect.\cite{wro}  

An exception to this pattern is the work of LeYouanc et al., who
have argued that inclusion of the $(70,1^-)$ intermediate states can
lead to a simultaneous resolution of the s/p-wave problem in ordinary
hyperon decay as well as the asymmetry problem in radiative hyperon
decay.\cite{LeY,LeY1}  In a previous work we have studied this 
approach within a chiral framework and have shown that it is indeed 
possible to find a simultaneous fit to both s- and p-wave hyperon 
decay amplitudes if contributions from the lowest lying ${1\over 2}^-$
and ${1\over 2}^+$ baryon octet resonant
states are included in the formalism.
\cite{BH2}  In the present paper we
extend this discussion to radiative hyperon decay.
In the next section then we introduce the 
effective weak and strong-electromagnetic Lagrangians including resonant
states and we evaluate the pole diagram contributions of both ground
state baryons and resonant states to radiative hyperon decay.
Numerical results are presented in Sec.~3,
while in Sec.~4 we conclude with a brief summary.
In the Appendix, we determine the strong couplings of the resonances
to the ground state octet by a fit to the electromagnetic decays of these
resonances.

\section{Radiative hyperon decays}
There exist six weak radiative hyperon decays:
$ \Sigma^+ \rightarrow p \gamma \, , \,
  \Xi^- \rightarrow \Sigma^- \gamma \, , \,
   \Sigma^0 \rightarrow n \gamma \, , \,
  \Lambda \rightarrow n \gamma \, , \,
  \Xi^0 \rightarrow \Sigma^0 \gamma \, , \,
  \Xi^0 \rightarrow \Lambda \gamma $
which can be studied experimentally, and each can be represented in
terms of a parity-conserving and a parity violating matrix element as
in Eq. (\ref{eq:bb}). The aim of the present work is to calculate the amplitudes
$A_{B'B}$ and $B_{B'B}$ within the framework of chiral perturbation theory and,
to this end, we consider first the Lagrangian
without resonances, which will be included in the following section.
It can be decomposed into a strong-electromagnetic and a weak part. 
The former reads
\beqa \label{ls}
{\cal L}^s_{B}  
 &=& 
 i \, \tr \Big( \bar{B} \gamma_{\mu} [  D^{\mu} , B] \Big) -
\mnod \, \tr \Big( \bar{B} B \Big) \no \\
&+ & l_d \, \tr \Big( \bar{B} \sigma_{\mu \nu}  \{ f^{\mu \nu}_+, B\} \Big)
+ l_f \, \tr \Big( \bar{B} \sigma_{\mu \nu} [ f^{\mu \nu}_+, B ] \Big) \qq ,
\eeqa
where $ f^{\mu \nu}_+ $ is the chiral field strength tensor of the
electromagnetic field and $\mnod$ represents the mass of the baryon octet
in the chiral limit. The coupling constants $l_d$ and $l_f$---so-called
low-energy-constants (LECs)---can be determined from a fit to the
baryon magnetic moments, {\it cf.} App.~A, and 
are the only counterterms contributing to the radiative hyperon 
decays up to second chiral
order. (Note, that at next chiral order,
there is an additional term possible which is proportional to
$\bar{B} \gamma_{\mu} [ D_{\nu} , f^{\mu \nu}_+] B$ but this vanishes 
for real photons.)

We now turn to the weak piece of the meson-baryon Lagrangian, whose
lowest order form is
\beq \label{lw}
{\cal L}_{\phi B}^{W}  =  \:
d \, \tr \Big( \bar{B}  \{ h_+ , B\} \Big) + \:
f \, \tr \Big( \bar{B}  [ h_+ , B ] \Big) \qquad .
\eeq
In our previous work the LECs $d$ and $f$ have been determined from 
the nonleptonic hyperon decays
by two independent means. In \cite{BH1} a calculation was performed
which included
{\it all} terms at one-loop order. This work suffered from the fact, however,
that at this order too many new unknown LECs enter the calculation so that
the theory lacks predictive power. In order to proceed these parameters 
were estimated  by means of spin-3/2 decuplet resonance exchange. The results
for the p-waves were still in disagreement with the data and,
therefore, additional counterterms that were not saturated by the
decuplet had to be taken into account, leading to the possibility of
an exact fit to the data.  In a second approach \cite{BH2} 
we included lowest lying spin-1/2$^+$ and 1/2$^-$ resonances
in the theory and performed a tree level calculation.
Integrating out the heavy degrees of freedom provides
then a plausible estimate of the weak counterterms, which
have been neglected completely in \cite{BSW} and \cite{J}.
A satisfactory fit for both s- and p-waves was achieved.
Since we herein apply this scheme for the weak radiative hyperon decays 
we will use the values for $d$ and $f$ from \cite{BH2}.

Having introduced the Lagrangian for the ground state baryons, we can then
proceed by including the low lying resonances.
In \cite{LeY,LeY1} it was argued that in a simple constituent 
quark model including the
lowest lying spin 1/2$^-$ octet from the (70,1$^-$) multiplet leads
to significant improvements in both radiative and nonleptonic hyperon 
decays. We confirmed in a recent calculation \cite{BH2} that indeed exist
significant contributions from these resonances for the nonleptonic hyperon
decays in the framework of chiral perturbation theory. 
We begin therefore with the inclusion of 
the octet of spin-parity 1/2$^-$ states, which include the
well-established 
states $N(1535)$ and $\Lambda(1405)$.  As for the rest
of the predicted 1/2$^-$ states there are a number of not so well-established
states in the same mass range---{\it cf.} \cite{LeY} and references therein.
In order to include resonances one begins by writing down the most
general Lagrangian at lowest order which exhibits the same symmetries
as the underlying theory, {\it i.e.} Lorentz invariance and chiral symmetry.
For the strong part we require invariance under $C$ and $P$ transformations
separately, while
the weak piece is invariant under $CPS$ transformations
where the transformation $S$
interchanges down and strange quarks in the Lagrangian.
We will work in the $CP$-conserving limit so that all LECs are real,
and denote the 1/2$^-$ octet by $R$. 

Under $CP$ transformations the fields behave as
\beqa
B \q & \rightarrow & \q \gamma_0 C \bar{B}^{T} \q , \q
\bar{B} \q \rightarrow  \q B^{T} C \gamma_0  \q , \q
f^{\mu \nu}_+ \q \rightarrow  \q - f_{\mu \nu +}^{T} \q , \q \no \\
h_+ \q & \rightarrow &  \q h_+^{T} \qq , \qq
D^{\mu} \q \rightarrow  \q - D_{\mu}^{T} \q , \q \no \\
R \q & \rightarrow & \q - \gamma_0 C \bar{R}^{T} \q , \q
\bar{R} \q \rightarrow  \q - R^{T} C \gamma_0  \q , \q
\eeqa
where $C$  is the usual charge conjugation matrix.
The kinetic term of the 1/2$^-$ Lagrangian is straightforward
\beq
{\cal L}_{R}^{kin}  
 = 
 i \, \tr \Big( \bar{R} \gamma_{\mu} [  D^{\mu} , R] \Big) -
M_R \, \tr \Big( \bar{R} R \Big)
\eeq
with $M_R$ being the mass of the resonance octet in the chiral limit.
The resonances R couple electromagnetically to the 1/2$^+$ baryon 
octet $B$ via the
Lagrangian
\beqa
{\cal L}^s_{RB}  
 &=& 
i r_d \, \Big[ \tr \Big( \bar{R} \sigma_{\mu \nu} \gamma_5
 \{ f^{\mu \nu}_+, B\} \Big)  + \tr \Big( \bar{B} \sigma_{\mu \nu} \gamma_5
 \{ f^{\mu \nu}_+, R\} \Big)  \Big] \no \\
& + &
i r_f \, \Big[ \tr \Big( \bar{R} \sigma_{\mu \nu} \gamma_5
 [ f^{\mu \nu}_+, B] \Big)  + \tr \Big( \bar{B} \sigma_{\mu \nu} \gamma_5
 [ f^{\mu \nu}_+, R] \Big)  \Big] \qq .
\eeqa
and the couplings $r_d$ and $r_f$ can be determined from a fit to the 
electromagnetic decays of the resonances---{\it cf.} App.~A.
The corresponding weak Lagrangian is
\beq
{\cal L}_{R B}^{W }  = 
i w_d \Big[ \, \tr \Big( \bar{R}  \{h_+ , B\} \Big)
       - \tr \Big( \bar{B}  \{ h_+, R\} \Big) \: \Big] +
  i w_f \Big[ \, \tr \Big( \bar{R}  [h_+ , B] \Big)
       - \tr \Big( \bar{B}  [ h_+, R] \Big) \: \Big] \q .
\eeq
with two couplings $w_d$ and $w_f$ which have been determined
from a fit to the nonleptonic hyperon decays in \cite{BH2}

We will not include additional resonances from the (70,1$^-$) states,
which were the {\it only} resonances considered in \cite{LeY,LeY1}. 
But in many applications the spin-3/2$^+$ decuplet and the 
Roper-like spin-1/2$^+$ octet states
play an important role, {\it cf.} {\it e.g.} \cite{BM}. The decuplet is only 
231 MeV higher in  average than the ground state octet and 
the Roper octet masses are comparable to the 1/2$^-$ states $R$.
One should therefore also account for these resonances.

Considering first the decuplet, due to angular momentum conservation
the spin-3/2 decuplet states can couple to the spin-1/2 baryon octet
only together with Goldstone bosons or photons.
Therefore, intermediate decuplet states can contribute only through
loop diagrams to radiative hyperon decay.
Such loop diagrams saturate
contact terms of the same chiral order as the loop corrections 
with the baryon octet.\cite{BH1}
Since in this work we restrict ourselves to lower chiral orders 
we can disregard such decuplet contributions. Their effect begins only
at higher chiral orders,
which we have neglected from the beginning.
In addition, the calculation of relativistic loop diagrams in the
resonance saturation scheme leads to some complications. 
The integrals are in 
general divergent and have to be renormalized which introduces
new unknown parameters. 
The absence of a strict chiral counting scheme in the relativistic 
formulation leads to contributions from higher loop diagrams
which are usually neglected in such calculations, {\it cf.} \cite{BM}.

Another important multiplet of excited states 
is the octet of
Roper-like spin-1/2$^+$ fields. While it was argued in \cite{JM} that
these play no role, a more recent study seems to indicate that one
cannot neglect contributions from such states to,
{\it e.g.}, decuplet magnetic moments.\cite{BaM} It is thus
important to investigate also the possible contribution of these baryon
resonances to the LECs.
The Roper octet, which we denote by $B^*$, 
consists of the $N^*(1440)$, the $\Sigma^*(1660)$, the 
$\Lambda^*(1600)$ and the $\Xi^*(1620?)$.
The transformation properties of $B^*$ under $CP$ are the same as for
the ground state baryons $B$, and
the effective Lagrangian of the $B^*$ octet coupled to the ground state
baryons takes the form
\beq
{\cal L}_{B^* B} = {\cal L}_{B^*}^{kin} + 
                   {\cal L}_{B^* B}^S + {\cal L}_{B^* B}^W
\eeq
with the kinetic term
\beq
{\cal L}_{B^*}^{kin} =
 i \, \tr \Big( \bar{B}^* \gamma_{\mu} [  D^{\mu} , B^*] \Big) -
M_{B^*} \, \tr \Big( \bar{B}^* B^* \Big) \q ,
\eeq
(with $M_{B^*}$ being the resonance mass in the chiral limit),
an electromagnetic interaction part 
\beqa
{\cal L}^s_{RB}  
 &=& 
  l_d^*\, \Big[ \tr \Big( \bar{R} \sigma_{\mu \nu} 
 \{ f^{\mu \nu}_+, B\} \Big)  + \tr \Big( \bar{B} \sigma_{\mu \nu} 
 \{ f^{\mu \nu}_+, R\} \Big)  \Big] \no \\
& + &
l_f^* \, \Big[ \tr \Big( \bar{R} \sigma_{\mu \nu} 
 [ f^{\mu \nu}_+, B] \Big)  + \tr \Big( \bar{B} \sigma_{\mu \nu}
 [ f^{\mu \nu}_+, R] \Big)  \Big] \qq .
\eeqa
and a weak piece
\beq
{\cal L}_{B^* B}^{W}  = 
d^* \Big[ \, \tr \Big( \bar{B}^*  \{h_+ , B\} \Big)
       + \tr \Big( \bar{B}  \{ h_+, B^* \} \Big) \: \Big] +
 f^* \Big[ \, \tr \Big( \bar{B}^*  [h_+ , B] \Big)
       + \tr \Big( \bar{B}  [ h_+, B^*] \Big) \: \Big] \q .
\eeq
As in the case of their 1/2$^-$ counterparts, the coupling constants $f^*,d^*$
have been be determined from nonleptonic hyperon decay in \cite{BH2},
while the electromagnetic couplings $l_d^*,l_f^*$ are found from
radiative decays of the resonances---({\it cf.} App.~A).
There exist no additional unknown parameters in this approach
once the weak couplings are fixed from the nonleptonic decays.
Study of radiative hyperon decay provides, therefore, a nontrivial
check on whether the results
from the simple quark model are consistent with chiral perturbation theory.

\subsection{Ground state contributions}
We begin by considering the diagrams which include only ground state
baryons, as depicted in Fig.~1.
The relevant Lagrangians are given in Eqs.~(\ref{ls}) and (\ref{lw}).
The photon couples not only via the field strength tensor but also through the
covariant derivative $D_{\mu}$
\beq
[D_{\mu},B] = \partial_{\mu} B + [ \Gamma_{\mu},B]
\eeq
Here the chiral connection
\beq
\Gamma_{\mu} = - i v_{\mu}  + \ldots = - i \, e \, Q \, A_{\mu} + \ldots
\eeq
contains the external photon field $A_{\mu}$ and the ellipses denote terms
that do not contribute in our calculation, while
the field strength tensor reads 
\beq
f_+^{\mu \nu} = 2 ( \partial^{\mu} v^{\nu} - \partial^{\nu} v^{\mu} ) 
+ \ldots \qq .
\eeq
The explicit calculation reveals that the contributions to the decays
from the chiral connection cancel if one 
uses the physical mass of the internal baryon for the propagator,
which is consistent to the order we are working.
Consequently, the only contribution to the radiative hyperon 
decays stems from the 
terms with the couplings magnetic $l_d$ and $l_f$.
Such pole diagrams with intermediate ground state baryons contribute only
to the parity conserving amplitudes $A^{ji}$, yielding\footnote{Corresponding
parity violating couplings vanish due to the Lee-Swift theorem.\cite{ls}} 
\beqa   \label{am}
A^{n \Sigma^0} &=& \frac{e}{M_\Sigma - M_N} \frac{ 8 \sqrt{2}}{3} \, d \, l_d 
\qq   \no \\
A^{n \Lambda} &=& \frac{e}{M_\Lambda - M_N} \frac{ 8 \sqrt{2}}{3\sqrt{3}} 
\, d \, l_d  \qq    \no \\
A^{\Sigma^0 \Xi^0} &=& \frac{e}{M_\Sigma - M_\Xi} \frac{ 8 \sqrt{2}}{3} 
\, d \, l_d  \qq   \no \\
A^{\Lambda \Xi^0} &=& \frac{e}{M_\Lambda - M_\Xi} 
\frac{ 8 \sqrt{2}}{3\sqrt{3}}  \, d \, l_d  \qq \no\\
A^{p\Sigma^+}&=&A^{\Sigma^-\Xi^-}=0
\eeqa
where we work in the limit of identical $u$- and $d$-quark
masses and employ the physical masses for the internal baryons.
If one were to include {\it only} ground state baryons in the effective theory
and neglect resonances then, there would be no additional contributions
to the amplitudes at tree level. This would lead to a vanishing asymmetry
parameter for {\it any} of the radiative hyperon decays and in
particular for the decay $\Sigma^+ \rightarrow p \gamma$.  
The inclusion of meson loops leads to a small contribution for
such asymmetry parameters\cite{N,JMLS} in clear contradiction to the
experimental result Eq. (\ref{eq:aa}).

\subsection{Resonance contributions}

Any explanation for the large asymmetry then must come from inclusion
of additional intermediate states.  
The diagrams including resonances are shown in Fig.~2. 
The spin-1/2$^-$
resonances contribute to the parity violating amplitudes
$B^{ji}$
\beqa   \label{amr}
B^{p \Sigma^+} &=& e \frac{4 (M_\Sigma - M_N)}{(M_\Sigma - M_R)(M_N - M_R)}
( \, \frac{1}{3} r_d + \, r_f ) ( \, w_d - \, w_f )
\qq ;  \no \\
B^{\Sigma^- \Xi^-} &=& e \frac{4 (M_\Xi -M_\Sigma)}{(M_\Sigma - M_R)
(M_\Xi - M_R)} ( \, \frac{1}{3} r_d - \, r_f ) ( \, w_d + \, w_f )
\qq ;  \no \\
B^{n \Sigma^0} &=& e \frac{4}{ (M_R -M_\Sigma)} \frac{\sqrt{2}}{3}
\, r_d  ( \, w_d - \, w_f )
+  e \frac{4}{ (M_R -M_N)} \frac{\sqrt{2}}{3}
\, r_d  ( \, w_d + \, w_f )
\qq ;  \no \\
B^{n \Lambda} &=& e \frac{4}{ (M_R -M_\Lambda)} \frac{\sqrt{2}}{3\sqrt{3}}
\, r_d  ( \, w_d + 3 \, w_f )
+  e \frac{4}{ (M_R -M_N)} \frac{\sqrt{2}}{3\sqrt{3}}
\, r_d  ( \, w_d - 3 \, w_f )
\qq ;  \no \\
B^{\Sigma^0 \Xi^0} &=& e \frac{4}{ (M_\Xi - M_R )} \frac{\sqrt{2}}{3}
\, r_d  ( \, w_d - \, w_f )
+  e \frac{4}{ (M_\Sigma - M_R)} \frac{\sqrt{2}}{3}
\, r_d  ( \, w_d + \, w_f )
\qq ;  \no \\
B^{ \Lambda \Xi^0} &=& e \frac{4}{ (M_\Xi - M_R )} \frac{\sqrt{2}}{3\sqrt{3}}
\, r_d  ( \, w_d + 3 \, w_f )
+  e \frac{4}{ (M_\Lambda - M_R)} \frac{\sqrt{2}}{3\sqrt{3}}
\, r_d  ( \, w_d - 3 \, w_f )
\qq .\label{eq:rr}
\eeqa
while the octet of spin-1/2$^+$
resonances contributes to the parity conserving amplitudes $A^{ji}$
\beqa   \label{amb}
A^{p \Sigma^+} &=& e \frac{4}{ M_\Sigma - M_{B^*}}
( \, \frac{1}{3} l_d^* + \, l_f^* ) ( \, d^* - f^* ) 
+ e \frac{4 }{M_N - M_{B^*}}
( \, \frac{1}{3} l_d^* + \, l_f^* ) ( \, d^* - f^* ) \qq ;  \no \\
A^{\Sigma^- \Xi^-} &=& e \frac{4}{ M_\Xi - M_{B^*}}
( \, \frac{1}{3} l_d^* - \, l_f^* ) ( \, d^* + f^* ) 
+ e \frac{4}{ M_\Sigma - M_{B^*}}
( \, \frac{1}{3} l_d^* - \, l_f^* ) ( \, d^* + f^* ) \qq ;  \no \\
A^{n \Sigma^0} &=& e \frac{4}{ M_\Sigma - M_{B^*}} \frac{\sqrt{2}}{3}
\, l_d^* ( \, d^* - f^* ) 
+ e \frac{4}{ M_{B^*}- M_N }\frac{\sqrt{2}}{3}
\, l_d^* ( \, d^* + f^* ) \qq ;  \no \\
A^{n \Lambda} &=& e \frac{4}{ M_\Lambda - M_{B^*}} \frac{\sqrt{2}}{3\sqrt{3}}
\, l_d^* ( \, d^* + 3 f^* ) 
+ e \frac{4}{ M_{B^*}- M_N }  \frac{\sqrt{2}}{3\sqrt{3}}
\, l_d^* ( \, d^* - 3 f^* ) \qq ;  \no \\
A^{\Sigma^0 \Xi^0} &=& e \frac{4}{ M_{B^*}-M_\Xi}  \frac{\sqrt{2}}{3}
\, l_d^* ( \, d^* - f^* ) 
+ e \frac{4}{ M_\Sigma - M_{B^*}}\frac{\sqrt{2}}{3}
\, l_d^* ( \, d^* + f^* ) \qq ;  \no \\
A^{\Lambda \Xi^0} &=& e \frac{4}{M_{B^*}- M_\Xi} \frac{\sqrt{2}}{3\sqrt{3}}
\, l_d^* ( \, d^* + 3 f^* ) 
+ e \frac{4}{ M_\Lambda - M_{B^*}}  \frac{\sqrt{2}}{3\sqrt{3}}
\, l_d^* ( \, d^* - 3 f^* ) \qq .
\eeqa
In the framework of chiral perturbation theory at tree level one {\it must}
include the spin-1/2$^+$ resonances in order to ensure a nonvanishing
asymmetry parameter for the decays $\Sigma^+ \rightarrow p \gamma$
and $\Xi^- \rightarrow n \gamma$ since the parity conserving component
vanishes if only the baryon octet is retained.   
In \cite{LeY1, JMLS} the coupling
of the photons to the ground state baryons was expressed directly in terms
of the experimental baryon magnetic moments which implicitly 
includes higher order chiral contributions and 
leads to nonvanishing parity conserving amplitudes for the
decays $\Sigma^+ \rightarrow p \gamma$
and $\Xi^- \rightarrow \Sigma^- \gamma$. As in the case of the nonleptonic 
hyperon decays one must to include the spin-1/2$^+$ resonances
to account for such higher order effects.\cite{BH2}

\section{Numerical results and discussion}
In this section we present the numerical results for the decay amplitudes
and the decay parameters.
For the electromagnetic couplings we use the values which can be obtained
from the magnetic moments of the ground state baryons
and the radiative decays of the resonances, {\it cf.} App.~A, while
the weak parameters are fixed from the nonleptonic hyperon decays.\cite{BH2}
For the parity conserving amplitudes we obtain, in units of 
$10^{-7}$ GeV$^{-1}$,
\beqa
A^{p \Sigma^+} &=&   0 - 1.81 = -1.81\qq \qq \qq 
A^{\Sigma^- \Xi^-} = \q 0 + 0.08 = 0.08    \no \\
A^{n \Sigma^0} &=&  0.50 - 0.52 = - 0.02  \qq  \qq \q \!  
A^{n \Lambda} \q =  \q 0.41 + 0.11 = - 0.52   \no \\
A^{\Sigma^0 \Xi^0} &=& -1.01 + 1.06 = 0.05 \qq \qq  \;  \:
A^{\Lambda \Xi^0} \q = \q  - 0.36 + 0.02 = -0.34 \qq ,
\eeqa
where the first and second numbers denote the contributions 
from the ground state octet and the spin-1/2$^+$ resonances, respectively.
The parity violating amplitudes read, in units of $10^{-7}$ GeV$^{-1}$,
\beqa
B^{p \Sigma^+} &=& 0.47  \qq \qq \q
B^{\Sigma^- \Xi^-} = \: 0.15     \no \\
B^{n \Sigma^0} &=& -0.45   \qq \qq \q
B^{n \Lambda} \; \, =  \:  -0.05 \no \\
B^{\Sigma^0 \Xi^0} &=& 0.70  \qq \qq \q \,
B^{\Lambda \Xi^0}  \;\;  = \; -0.08 \qq .
\eeqa
The decay rates are given by
\beq
\Gamma^{ji}  = \frac{1}{\pi} \Big( \frac{M_i^2 -M_j^2}{ 2M_i} \Big)^3
    \bigg( |A^{ji}|^2 + |B^{ji}|^2  \bigg) 
\eeq
and one obtains, in units of  GeV
\beqa
\Gamma^{p \Sigma^+} &=&  1.3 \: 10^{-16}  \q ( 1.0 \, 10^{-17} ) \qq  \qq 
\Gamma^{\Sigma^- \Xi^-} =  \q 1.6 \: 10^{-19}  \q ( 5.3 \, 10^{-19} )  \no \\
\Gamma^{n \Sigma^0} &=&  7.5 \: 10^{-18} \qq  \qq \qq  \qq \q \; \: 
\Gamma^{n \Lambda}  \q \: = \q  3.8 \: 10^{-18} \q ( 4.6 \, 10^{-18} )\qq\no \\
\Gamma^{\Sigma^0 \Xi^0} &=& 2.7 \: 10^{-18}  \q ( 7.9 \, 10^{-18} ) \qq  \qq 
\Gamma^{\Lambda \Xi^0} \q \! = \q  2.5 \: 10^{-18}  \q ( 2.5 \, 10^{-18} )\qq ,
\eeqa
where the number in the brackets denotes the experimental value.
(The decay $\Sigma^0 \rightarrow n \gamma$ is dominated by the
electromagnetic decay $\Sigma^0 \rightarrow \Lambda \gamma$ and, therefore,
no experimental value can be given in this case.)  Finally, the
corresponding asymmetry parameters are found to be, {\it cf.} Eq.~(\ref{eq:qq}),
\beqa
\alpha^{p\Sigma^+}  &=& -0.49  \qq  \qq 
\alpha^{\Sigma^- \Xi^-} =  \q 0.84 \no \\
\alpha^{n \Sigma^0} &=&  0.12 \qq  \qq \q 
\alpha^{n \Lambda} \q  =   \q -0.19 \no \\
\alpha^{\Sigma^0 \Xi^0} &=& 0.15 \qq  \qq \q 
\alpha^{\Lambda \Xi^0} \; \;  =  \q 0.46 \qq .
\eeqa

Our results are only indicative, of course.  A full discussion would
have to include both the effects of chiral loops as well as 
contributions from additional resonant states.  In this regard, we do not
anticipate that our predictions should be able to precisely reproduce
the experimental values for the decay widths, but it should be noted
that we obtain in our approach radiative hyperon decay widths which
are in reasonable agreement with experiment except in the case 
of $\Sigma^+ \rightarrow p \gamma$, which is about an order of magnitude larger
than the experimental value. We could, of course, 
by use of average resonant mass
rather than non-strange mass or by twiddling parameters, bring this
number into better agreement with experiment.  However, our purpose 
herein is not a precise fit
to data, but rather to ask whether the resonance saturation is able to
represent the basic phenomenology of these decay rates.  
In this regard, the answer is then yes---our results suggest that 
the spin-1/2$^-$ resonances
play an {\it essential} role for the radiative decays as was found 
in the constituent quark model.\cite{LeY1} (Indeed at lowest chiral
order without such resonant contributions both parity conserving and
parity violating amplitudes would vanish!)  But in addition, in the
chiral approach one {\it must}
include the octet of spin-1/2$^+$ resonances in order to account for
SU(3) breaking effects, which are higher order in the chiral
expansion.  A similar conclusion was reached in the case of the 
nonleptonic hyperon decays.\cite{BH2}

What {\it is} perhaps more important here is that with the inclusion 
of resonant
contributions, the origin of the ``large'' negative asymmetry in the radiative
$\Sigma^+$ hyperon decay is no longer a mystery.  Indeed it becomes
almost natural.  The resonant contribution to the parity conserving
and parity violating amplitudes are comparable in size, leading to
significant asymmetries for a number of the radiative modes, including
$\Sigma^+\rightarrow p\gamma$.  It should also be noted that there is
no conflict with Hara's theorem.  Examination of Eq. (\ref{eq:rr})
clearly shows that the parity violating amplitudes for the decays
$\Sigma^+\rightarrow p\gamma$ and $\Xi^-\rightarrow \Sigma^-\gamma$
vanish in the SU(3) limit.

\section{Summary}
In this work we examined the significance of low
lying baryon resonant contributions to radiative hyperon decay. 
To this end, we included the spin-1/2$^-$ octet from the (70,1$^-$) states
and the octet of Roper-like 1/2$^+$ fields in the
effective theory. 
The most general Lagrangian incorporating these resonances coupled to the 
ground state baryons introduces twelve new parameters, four of which can be
determined from the electromagnetic decays of the resonances,
two can be fitted from the ground state baryon magnetic moments and
the remaining six weak couplings
have already been determined from nonleptonic hyperon
decays within the framework of chiral perturbation theory.\cite{BH2}
Thus, the inclusion of the spin-1/2 resonant states leads to
no additional unknown parameters.  (It should be noted that an
alternative approach---inclusion of the spin-3/2$^+$ decuplet, as performed
in \cite{BH1} for nonleptonic hyperon decay, generates terms at the
same chiral order as the loop corrections---{\cal O}($p^2$), 
which is beyond the accuracy
of this calculation and therefore can be neglected.)
In \cite{LeY,LeY1} it was argued that within the quark model 
the inclusion of the spin-1/2$^-$ octet
is sufficient to obtain a satisfactory fit for both nonleptonic
and radiative hyperon decays.
We have shown that in the framework of chiral
perturbation theory the structure of the contributions from such resonances
agrees with the results in the quark model {\it to the order we are working}.
In \cite{LeY1} the pole terms of the ground state baryons
to the parity conserving amplitudes
were expressed in terms of {\it experimental} magnetic moments and
thereby a significant nonzero value for the
asymmetry parameter in polarized $\Sigma^+ \rightarrow p \gamma$ was obtained.
However, this approach includes the anomalous magnetic moment components which
are of higher chiral order and, therefore, do not appear to the order
we are working.  On the other hand, in our tree level chiral
perturbative calculation
the improvement of experimental agreement is brought about by
the inclusion of the Roper-octet, which is in the same mass
range as the 1/2$^-$ octet.  We found that
by using the electromagnetic couplings determined from a fit to the
magnetic moments of the ground state baryons and from the electromagnetic
decays of the resonances, and the weak couplings from the nonleptonic
hyperon decays 
we obtained reasonable predictions for the decay amplitudes and
significant negative values for the $\Sigma^+\rightarrow p\gamma$ 
asymmetry as a very
{\it natural} result of this picture, even though Hara's theorem is satisfied.

We conclude that the inclusion of spin-1/2 resonances in nonleptonic
hyperon decays provides a reasonable explanation
of the importance of higher order 
counterterms and gives a satisfactory picture of both radiative and
nonradiative nonleptonic hyperon decay.  In order to make a more 
definite statement one should, of course, go to higher orders 
and include meson loops as well as the contributions from 
additional resonances.  However, this
is clearly beyond the scope of the present investigation.

\section*{Acknowledgements}
This work was supported in part by the Deutsche Forschungsgemeinschaft
and by the National Science Foundation.

\appendix 
\def\theequation{\Alph{section}.\arabic{equation}}
\setcounter{equation}{0}
\section{Determination of the strong couplings} \label{app:a}

In this appendix we present the determination of the photon baryon couplings
used in the effective Lagrangian. We start with the ground state
baryons.
In this case the two appearing LECs $l_d$ and $l_f$ can be fit
to the baryon magnetic moments which are defined by
\beq
\mu^{ji} = \frac{e}{M_i + M_j} \, \Big[ \, F_1^{ji}(0) + \, F_2^{ji}(0) \Big]
= \frac{e}{M_i + M_j} \, F_1^{ji}(0) + \, A^{ji}
\eeq
The form factor $F_1^{ji}(0)$ is related to the electromagnetic
charges of the baryons
\beq
F_1^{ji}(0) = \, q_i \, \delta_{ji}    \qq , \qq q_i = \{-1,0,1\} \qq.
\eeq
A fit for $l_d$ to the magnetic moments of the ground state baryons delivers
\beq
l_d = 0.25 \: \mbox{GeV}^{-1}
\eeq
and we neglected the uncertainties in our fit since we are only interested
in the order of magnitude for this parameter. (Note, that $l_f$ does
not contribute to the amplitudes at tree level.)

We now turn to the determination of the couplings $r_d, r_f$ and $l_d^*,l_f^*$ 
appearing in the electromagnetic part of the effective
resonance-ground state Lagrangian.
The decays listed in the particle data book,
which determine the coupling constants $r_d$ and $r_f$,
are $N(1535) \rightarrow N \gamma$.
The width is given by 
\beq  \label{dec}
\Gamma^{ji} = \frac{1}{8 \pi M_R^2} |{\bf k}_\gamma| |{\cal T}^{ji}|^2
\eeq
with
\beq  \label{ene}
|{\bf k}_\gamma| = E_\gamma = \frac{1}{2 M_R} (M_R^2- M_B^2) 
\eeq
being the three-momentum of the photon 
in the rest frame of the resonance and $M_R$ and $M_B$ being the
masses of the resonance and the
ground state baryon, respectively.
For the resonance mass $M_R$ we use the mass of $N(1535)$.
The mistake we make in the case the other resonance octet states
is of higher chiral order and, therefore, beyond the accuracy
of our calculation.
For the transition matrix one obtains
\beq
|{\cal T}^{ji}|^2 = \, 128 \, e^2 \, ( p_i \cdot k)^2 \, (C^{ji})^2
\eeq
with $p_i$ the momentum of the decaying baryon and the coefficients
\beq
C^{p \,p(1535)} = \frac{1}{3} \, r_d + \, r_f   \qq , \qq
C^{n \,n(1535)} = - \frac{2}{3} \, r_d   \qq .
\eeq
Using the experimental values for the decay widths
we arrive at the central values
\beq
e \, r_d = 0.033 \: \mbox{GeV}^{-1} \qq , \qq 
e \, r_f =   - 0.046 \: \mbox{GeV}^{-1} \qq .
\eeq
We do not present the uncertainties in these parameters here, since
for the purpose of our considerations a rough estimate of these constants
is sufficient.

For the determination of $l_d^*$ and $l_f^*$ we use the decays
$N(1440) \rightarrow N \gamma$.
One has to replace the resonance mass by $M_{B^*} \simeq 1440$ MeV 
in Eq.~(\ref{dec},\ref{ene}) and the coefficients read
\beq
C^{p \,p(1440)} = \, \frac{1}{3} \, l_d^* + \, l_f^*   \qq , \qq
C^{n \,n(1440)} = - \frac{2}{3} \, l_d^*   \qq .
\eeq
The fit to the decay widths delivers
\beq
e \, l_d^* = - 0.024 \: \mbox{GeV}^{-1}  \qq , \qq 
e \, l_f^* = - 0.009 \:\mbox{GeV}^{-1}  \qq .
\eeq

\newpage


\section*{Figure captions}

\begin{enumerate}

\item[Fig.1] Diagrams that contribute to radiative hyperon decays.
             Solid and wavy lines denote ground state baryons
             and photons, respectively.
             Solid squares and circles are vertices of the
             weak and electromagnetic interactions, respectively.

\item[Fig.2] Diagrams including resonances that contribute to radiative hyperon
             decays. Solid and wavy lines denote ground state baryons
             and photons, respectively. The double line represents
             a resonance. Solid squares and circles are vertices of the
             weak and electromagnetic interactions, respectively.

\newpage

\begin{center}
 
\begin{figure}[bth]
\centering
\centerline{
\epsfbox{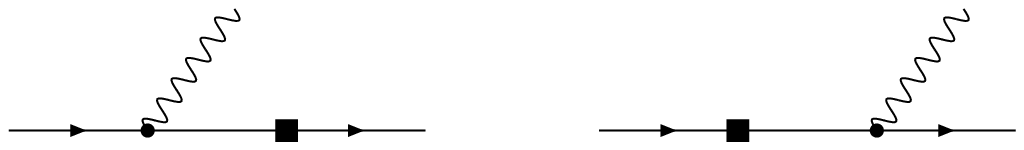}}
\end{figure}

\vskip 0.7cm

Figure 1

\vskip 1.5cm

\begin{figure}[tbh]
\centering
\centerline{
\epsfbox{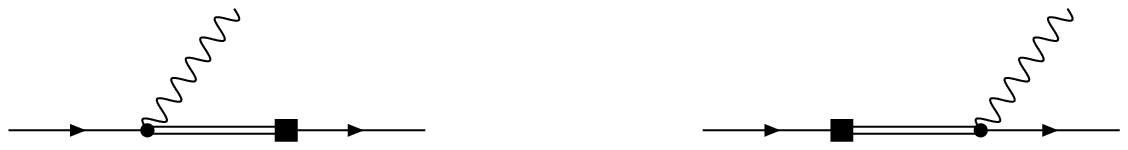}}
\end{figure}

\vskip 0.7cm

Figure 2

\end{center}

\end{enumerate}

\end{document}